\documentclass[12pt,preprint]{aastex}
\begin{document}
\title{Progenitors of Type Ia Supernovae: \\ Binary Stars with White
 Dwarf Companions} \author{M. Parthasarathy, David
 Branch, David~J. Jeffery, \& E.~Baron} \affil{Homer. L. Dodge
 Department of Physics and Astronomy, University of Oklahoma, Norman,
 Oklahoma 73019, USA} 

\begin{abstract}

Type~Ia SNe (SNe~Ia) are thought to come from carbon--oxygen white
dwarfs that accrete mass from binary companions until they approach
the Chandrasekhar limit, ignite carbon, and undergo complete
thermonuclear disruption.  A survey of the observed types of binaries
that contain white dwarfs is presented.  We propose that certain
systems that seem most promising as SN~Ia progenitors should be more
intensively observed and modeled, to determine whether the white
dwarfs in these systems will be able to reach the Chandrasekhar limit.
In view of the number of promising single--degenerate systems and the
dearth of promising double--degenerate systems, we suspect that
single--degenerates produce most or perhaps all SNe~Ia, while
double--degenerates produce some or perhaps none.

\end{abstract}


\section{INTRODUCTION}

Type~Ia supernovae (SNe~Ia) used as distance indicators for cosmology
have revealed that the cosmic expansion is accelerating owing to the
existence of some kind of ``dark energy'', and plans are being made to
discover and carefully observe numerous high--redshift SNe~Ia in order
to probe the nature of the dark energy.  Thus it is important to
identify the stellar progenitors of SNe~Ia.  The exploding star is
thought to be a carbon--oxygen (CO) white dwarf (WD) that accretes
mass from a binary companion until it approaches the Chandrasekhar
limit (CL) of 1.4~$M_\odot$, ignites carbon under electron--degenerate
conditions, undergoes a thermonuclear instability, and disrupts
completely (Nomoto et~al. 1984; Woosley \& Weaver 1986; Leibundgut
2000, 2001; Hillebrandt \& Niemeyer 2000).  However, the nature of the
mass--donating star is not yet known (Branch et~al. 1995).

About 75 percent of the observed WDs have masses near 0.6~M$_{\odot}$,
about 10 percent have lower mass near 0.4~M$_{\odot}$, and about 15
percent have higher mass above about 0.8~M$_{\odot}$ (Liebert et~al.
2005a; Vennes 1999).  The SN~Ia progenitor WD therefore must be a
member of a close binary in which it can accrete at least a few tenths
of a solar mass in order to reach the CL.

Two main scenarios have been proposed, involving either the merger of
two WDs (the double--degenerate, or DD, scenario; Iben and Tutukov
1984), or a single WD that accretes from a normal companion star (the
single--degenerate, or SD, scenario; Whelan and Iben 1973). Some
population synthesis studies favor the DD scenario (e.g., Yungelson \&
Livio 1998), and recent observational studies have identified many DD
binaries (Napiwotzki et~al. 2004).  However, not one of the observed
DD systems has both an orbital period short enough to merge in a
Hubble time and a total mass that exceeds the CL.  In addition, the
merger of two CO WDs may lead to accretion--induced collapse to a
neutron star rather than to a SN Ia (Timmes et~al. 1994; Nomoto \&
Iben 1985; Mochovitch et~al. 1997; but see Piersanti et~al. 2003).
Therefore, the SD scenario is generally favored today. The most recent
theoretical models of accreting WDs in the SD scenario find that some
of them can reach the CL (Langer et~al. 2000; Yoon \& Langer 2003; Han
\& Podsiadlowski 2004).

In addition to population synthesis studies and theoretical models of
accreting WDs, a more observational approach can be helpful. In this
paper we consider various types of observed binaries that contain a WD
(the SD scenario), and consider which of them are good candidates for
approaching the CL.

The observed companions of WDs in binaries include main--sequence
stars (MS), red--giant (RGB) stars, asymptotic--giant--branch (AGB)
stars, and post--AGB stars.  About 20 stars in the Yale Bright Star
Catalog are known to have WD companions. This is an underestimate of
the true number, in view of the observational difficulty of detecting
WD companions of bright stars of spectral type earlier than F.  The
WDs can be spatially resolved in classical nearby visual binaries such
as Sirius and Procyon, but in unresolved systems the WD is difficult
to detect by optical spectroscopy. The WD can be detected at shorter
wavelengths, provided that it is hotter than the primary star.

\section{WHITE DWARF + MAIN--SEQUENCE STAR BINARIES}

Barstow et~al. (2001) resolved several Sirius--like binaries with a
Hubble Space Telescope (HST) UV imaging survey of stars known to have
hot WD companions that are unresolved from the ground.  Of the 17
systems observed, eight were resolved with the Wide Field Planetary
Camera~2, using various UV filters. Some of the unresolved systems
seem to be close binaries such as HR~8210 = IK~Peg (A8V), which has a
short orbital period of 21.7 days (Vennes et~al. 1998). Barstow
et~al. found 56~Per (F4~V), whose WD companion has a mass of 0.9
M$_{\odot}$ (Landsman et~al.  1996), and 14~Aur (F4~V) to be quadruple
and quintuple, respectively. Close binary systems with WDs and
multiple companions are important for understanding the dynamical,
accretion, and evolutionary processes of WDs in such systems.
 
The ROSAT WFC and EUVE sky surveys have produced samples of WDs with
K~IV-V, G~IV-V, F~IV-V, A~III-V, and low--mass M~V secondary
stars. Many of these systems show orbital characteristics of
post--common--envelope (post--CE) binaries (Hillwig et~al. 2000). A
few systems composed of a DA~WD and secondary stars with spectral
types B to K and luminosity classes from V to III were found in IUE
low resolution UV spectra. For example, HR~8210 contains an A8~V star
and a DA WD with a mass well in excess of 1.0~M$_{\odot}$. Vennes et
al. (1998) studied some of the hot WDs in the EUVE Survey. Parameters
of selected binaries with WD and MS stars of spectral types B1.5~V to
K~V are given in Table~1.

\begin{deluxetable}{rrrrr}
\tablecolumns{5} \tablewidth{0pc} \tablecaption{SELECTED WHITE DWARF +
MAIN SEQUENCE BINARIES} \tablehead{ \colhead{}& \colhead{V} &
\colhead{Sp} & \colhead{WD (M$_{\odot}$)} & \colhead{Period (days)}}
\startdata HR~8210 = IK~Peg = EUVE J2126+193 & 6.07 & A8 V & 1.25 &
21.72 \\ HR 2875 (triple system) & 5.42 & B3.5 V+B6 V & 1.0 & 15.081
\\ HD 209295 & 7.3 & A9 V & 1.04 & 3.106 \\ 16 Dra (triple) & 5.51 &
B9 V & 0.8 & \nodata \\ HD 33959C = EUVE J0515+326 (triple)& 7.95 &
F4IV--V & 0.7 & \nodata \\ EUVE J0702+129 & 10.0 & K0 V & 0.93 &
\nodata \\ EUVE J0228-613 = HD 15638 & 8.8 & F6 V & 1.10 & \nodata \\
HD 223816 = EUVE J2353-703 & 8.8 & G0 V & 0.95 & \nodata \\ HD 217411
= EUVE J2300-070 & 9.8 & G5 V & 1.16 & \nodata \\ EUVE J0044+095 = BD
+08 102& 10.16 & K2 V & 0.9 & \nodata \\ EUVE J1925-565 & 10.6 & G 5 V
& 0.76 & \nodata \\ EUVE J1024+263 = HD 90052 & 9.6 & F0 V & 0.93 &
\nodata \\ EUVE J0357+286 & 11.7 & K 2 V & \nodata & \nodata \\ EUVE
J0356-366 & 12.45 & G2 V & \nodata & \nodata \\ EUVE J1027+323 & 13.0
& G5 V & \nodata & \nodata \\ \enddata
\end{deluxetable}

Burleigh \& Barstow (2000) found that the B9~V star 16~Dra has a WD
companion. White--dwarf companions to B and A stars are important
since they provide an observational lower limit on the maximum mass of
a MS star that can become a WD. Since WDs in such systems are expected
to be massive (0.8 to 1.0 M$_{\odot}$) and the MS stars evolve
rapidly, WDs in such close binaries may evolve to SNe~Ia. Handler
et~al. (2002) found that the single--lined eccentric--orbit
spectroscopic close binary HD~209295, which has a 3.1~day orbital
period, consists of an A9~V primary and a 1.04~M$_{\odot}$ WD.  The
system seems to be similar to HR~8210. Vennes
(2000) found that HR~2875 is a close triple system with B3.5~V + B6~V
+ a WD in excess of 1.0~M$_{\odot}$. The B3.5~V + B6~V stars have an
orbital period of 15~days with orbital eccentricity of 0.68. The close
binary WD in HR~2875 is detected in the EUVE spectrum (Burleigh \&
Barstow 1998); its $T_\mathrm{eff}$ is estimated to be $\sim$45,000~K. The
WDs in HR~2875--like systems are expected to be massive since the MS
companions are B3~V to B5~V stars.  The GALEX UV survey may reveal
additional close binaries with late B and early A stars plus hot and
massive WDs, which will be candidate progenitors of SNe~Ia.

There are several hot WD + M--dwarf binaries (Green et~al. 2000), but
the total mass of most such systems is much less than the CL.  In any
case, since the evolution of M dwarfs is very slow most of them will
not become cataclysmic variables or SNe~Ia. Frequently, the WDs in the
Sloan Digital Sky Survey (SDSS) are accompanied by an unresolved or
barely resolved MS (nearly always M dwarf) companion with a composite
spectrum. Raymond et~al. (2003) studied 109 of these in more detail,
with the main goal of finding close pairs that might be
pre-cataclysmic variables. They found that the WDs in these systems
are fairly hot, with $T_\mathrm{eff}$ from 8000 to 42,000~K. However in all
these systems the companion is an M dwarf. With the release of the
third SDSS data set, some 501 such pairs have been discovered in which
the companions are all M dwarfs (Silvestri et~al. 2005).

From the analysis of the extensive SDSS data it may be possible to
find composite spectra of hot WDs and late B, A and F stars.

Another category of binaries with WD + MS stars is that of dwarf
carbon stars (Green et~al. 2000).  These dwarfs became carbon--rich as
a result of mass transfer during the AGB phase of the present WD
companions.  Some of the dwarf carbon stars with very high carbon
abundances are expected to have hot WD companions (Heber et~al. 1993;
Liebert et~al. 1994).  Further observational study of a large sample
of dwarf carbon stars is needed.

%\end{document}

\subsection{V~471 TAURI--TYPE BINARIES WITH WHITE--DWARF COMPANIONS}

V~471 Tauri is a remarkable system in the Hyades cluster.  It is an
eclipsing binary whose components are a hot DA WD of 0.84~M$_{\odot}$
and a K2 MS star of 0.93~M$_{\odot}$, with an orbital period of only
12.5 hours. The high effective temperature and high mass of the WD
present an evolutionary paradox (O'Brien et~al. 2001). The WD is the
most massive one known in the Hyades, but also the hottest and
youngest, in direct conflict with expectation.

O'Brien et~al. (2001) conclude that the WD is descended from a blue
straggler. They suggest that the progenitor system was a triple. Two
of the components merged, leaving a blue straggler and the K2~V
star. The blue straggler evolved to the AGB phase and common envelope
interaction with the K2~V star, which spiraled in to its present
separation and ejected the envelope. The present K2~V component is
found to be 18 percent oversized for its mass relative to normal
Hyades K2~V stars of same mass. The evolution of the orbital period
and the spiraling in of the K2~V star need to be explored to see if
V~471 Tau--type systems can result in SNe~Ia.  Parameters of selected
V~471 Tau--type binaries are given in Table~2.

\begin{deluxetable}{rrrrr}
\tablecolumns{5}
\tablewidth{0pc}
\tablecaption{V 471 TAU--TYPE CLOSE BINARIES}
\tablehead{\colhead{}&\colhead{Period (days)}&\colhead{WD (M$_{\odot}$)}
& \colhead{$T_\mathrm{eff}$}&\colhead{K V (M$_{\odot}$)}}
\startdata
 V 471 Tau & 0.5212 & 0.84 & 34,500 K & 0.93 \\
 HS 1136+6646& 1 & \nodata & 100,000 K & \nodata \\
 BE UMa & 2.291 & 0.7 & 105,000 K & \nodata \\
 EC 13471 - 1258 & 0.151 & 0.77 & 14,085 K& 0.58 \\ 
\enddata
\end{deluxetable}

Common--envelope evolution is not well understood. However several
types of binaries, such as cataclysmic variables, SN~Ia progenitors,
X-ray binaries, etc., seem to go through the CE phase. Post-CE
binaries provide important observational constraints on theoretical
models. Only about 25 post--CE binaries have measured orbital periods
and far fewer have accurate component masses (Hillwig
et~al. 2000). The post--CE binaries BE~UMa (sdO/DAO + K5~V; 2.29~day
orbital period) and HS~1136+6646 (DAO + K7~V; 0.836~day orbital
period) appear to be similar to V~471 Tau, except that BE~UMa and
HS~1136+6646 have very hot (105,000~K and 70,000~K) WDs. In all three
systems the K~V star is out of thermal equilibrium, resulting in radii
larger than normal. The total mass of the components in BE~UMa
probably is close to the CL, as in V~471 Tau.  Hillwig
et~al. (2000) listed several additional post--CE systems with very hot
WD companions, but most of these have longer orbital periods and their
total masses seem to be much less than the CL.

Schreiber \& Gansicke (2003) analysed 30 well--observed post-CE
binaries. In their sample only V~471 Tau has a total mass above the
CL, although that of EC 13471-1258 (orbital period 0.151~days; WD mass
of 0.77~M$_{\odot}$; and mass of the MS component 0.58 M$_{\odot}$) is
very close to it. Schreiber \& Gansicke calculated the orbital
evolution of the systems and concluded that they will evolve into
semi--detached configurations and begin mass transfer. They predict
that the orbital period of V~471 Tau will decrease from its present
0.521~days to 0.167~days. Therefore V~471 Tau type systems can be
considered as candidate progenitors of SNe~Ia.

The GALEX UV survey may reveal a larger sample of binaries with hot,
massive WDs and MS (and RGB, AGB, and post--AGB) companions.  From a
study of these systems we may be able to find some with orbital
periods and accretion rates that are appropriate for evolution into
SNe~Ia.
      
%\end{document}

\section{BINARY ORIGIN OF MASSIVE WHITE DWARFS}
   
Vennes (1999) and Marsh et~al. (1997a,b) studied the properties of hot
WDs in the EUVE and ROSAT surveys and found several massive WDs.
Vennes (1999) listed 15 WDs more massive than 1~M$_\odot$, and found
in these a high incidence of magnetic WDs. He suggested that the
magnetism may be due to magnetism of the MS progenitors and/or to
post--AGB evolution affected by magnetic fields. He also suggested
that the massive WDs may be products of close binary evolution.

Liebert et~al. (2005a) studied 348 DA~WDs in the Palomar Green (PG)
survey and found eight high--mass WDs in their sample, whereas less
than one was expected on the basis of the standard WD mass
distribution.  The average mass of the eight high--mass WDs is
0.93~M$_{\odot}$, including PG~1658+441 (1.31~M$_{\odot}$; Schmidt
et~al. 1992).  In view of the apparent excess of high--mass WDs,
several authors have suggested that a substantial fraction of the 0.8
to 1.35~M$_{\odot}$ WDs may result from close binary evolution, i.e.,
mergers of helium + helium~WDs, helium~WD + CO~WD, and CO + CO~WDs.

If some of these massive WDs have close companions, accretion
processes or mergers might produce SNe~Ia, in which case some of the
SN~Ia progenitors are initially very close triple systems (in addition
to the triple systems suggested for the origin of V~471 Tau~type
binaries discussed above).

From Sloan Digital Sky Survey data Kleinman et~al. (2004) found 2500
new hot WDs, including 20 of high mass. Also several magnetic WDs have
been found (Gansicke et~al. 2002; Schmidt et~al. 2003). A total of 169
magnetic WDs are now known, most having fields higher than
2~MG. Magnetic WDs with fields higher than 1~MG constitute about 10
percent of all WDs and have a mean mass of 0.93~M$_{\odot}$ compared
to 0.56~M$_{\odot}$ for all WDs. Liebert et~al. (2005b) noticed a
curious, unexpected property of the total lists of magnetic WDs and of
WD + MS binaries: there appears to be virtually zero overlap between
the two samples. No confirmed magnetic WD has been found in a system
with a MS star. This contrasts with the situation for interacting
binaries, in which an estimated 25 percent of the accreting systems
have a magnetic WD. It is possible that some of the magnetic WDs may
have very close, unresolved MS companions of spectral type earlier
than M, as opposed to the non--magnetic WDs in the SDSS data in which
the companion is nearly always an M~dwarf. It may be a selection
effect in view of the large mass and small radius of magnetic WDs
(Liebert et~al. 2005a,b). The magnetic WDs have a higher than average
mass because on average they have more massive progenitors, rather
than because the initial--final mass relation (IFMR) is affected by
the magnetic field of the progenitors (Wickramasinghe \& Ferrario
2005). The magnetic WD progenitors may be stars with initial mass more
than 4.5~M$_{\odot}$ that had magnetic fields during the MS phase or
generated magnetic fields during post--MS phases (Wickramasinghe \&
Ferrario 2005). Since a significant fraction of MS stars more massive
than 4.5~M$_{\odot}$ are close binaries, it is expected that a
fraction of the magnetic WDs may contain unresolved companions. Radial
velocity monitoring and multiwavelength observations of magnetic WDs
are needed to search for companion stars and to understand the
evolution of such systems into SNe Ia.
   
The IFMR is constrained by observations of WDs in clusters (Weidemann
2000). There is not much observational information on the IFMR at the
high mass end.  The current best estimate for the maximum initial mass
is 6.5~M$_{\odot}$, producing a WD of about 1~M$_{\odot}$.  Williams
et~al. (2004) and Kalirai et~al. (2005) constrain the upper mass limit
of WD progenitors to be 5.8~M$_{\odot}$ for a cluster age of
150~Myr. There seem to be no magnetic WDs in young open clusters, so
the effect of magnetic field on the IFMR is unknown.

%\end{document}

\section{F, G, AND K GIANTS WITH WHITE--DWARF COMPANIONS}

In addition to MS stars with WD companions, there are F, G, and K
giants with WDs. The Ba~II G~III and K~III stars and the CH stars are
all have WD companions (McClure 1984; McClure \& Woodsworth 1990;
B\"ohm--Vitense et~al. 2000), and carbon, barium, strontium, and other
s--process heavy elements are overabundant due to mass accretion from
the former AGB companion (now the WD). If Ba~II and CH stars are
assumed to have masses of 1.5~M$_{\odot}$ and 0.8~M$_{\odot}$,
respectively, then the mass functions derived from the spectroscopic
orbits are such that the WD companions are around 0.6~M$_{\odot}$
similar to that of a typical field WD. The WD companions of a few
Ba~II stars with short orbital periods may accrete enough mass to
reach the CL during the Roche lobe overflow or during the AGB phase of
the present G--K giants, as the total mass of the some Ba~II binaries
is expected to be more than 2~M$_{\odot}$.  The total masses of
CH--star systems appear to be too low to produce SNe~Ia.

Based on UV spectra of Ba~II stars obtained with the HST,
B\"ohm--Vitense et~al. (2000) concluded that it is indeed highly
probable that all Ba~II and mild Ba~II stars have WD companions, which
in most cases are rather cool and therefore old.  Most Ba~II stars
must have come from B or possibly early A MS stars.  The MS F stars,
with masses around 1.5~M$_{\odot}$, do not produce many Ba~II stars.
The incidence of Ba~II stars among giants is about one percent
(B\"ohm-Vitense et~al. 2000).  Therefore it is expected that nearly
one percent of MS early A and late B have WD companions, which are
expected to be hot and young. Systematic radial--velocity and far--UV
surveys are needed to search for such systems, which may be candidate
SN~Ia progenitors.

Subgiant CH stars (Bond 1974; Luck \& Bond 1991; McClure 1997), dwarf
Ba~II stars (North et~al. 2000), dwarf carbon stars, and F~V stars
with strong Sr~II 4077~\AA\ lines (North et~al. 2000) also have WD
companions and overabundances of carbon and s--process elements. It is
likely that subgiant CH stars and dwarf Ba~II stars are precursors to
giant Ba~II stars.  Most of the known Ba~II, CH, and subgiant CH
stars, etc., are very long period binaries (McClure \& Woodsworth
1990; McClure 1997) and the masses of the primaries are low, therefore
most of these systems are not candidate SN~Ia progenitors.

The Tc--poor S stars are evolved Ba~II stars (Iben \& Renzini 1983)
with orbital elements similar to Ba II stars. The no--Tc S stars
consist of mass--losing S giants (on the AGB) with WD companions
(Jorissen 1999; Van Eck \& Jorissen 2002). Orbital periods are found
to be similar to those of symbiotic systems.  Some no--Tc S--star
binaries have been found to show symbiotic activity (Van Eck \&
Jorissen 2002). Symbiotic binary S stars are more evolved than
non--symbiotic binary S stars, and mass-loss from the S star and
accretion onto the WD produces the symbiotic activity.

Further studies, especially far UV, of Ba II G--K giants, CH giants,
subgiant CH stars, F~V stars with strong Sr~II 4077~\AA\ lines, dwarf
carbon stars, no--Tc S stars, and yellow symbiotics may reveal some
short--period systems with massive WD companions. We do not have
orbital parameters for many of these systems.

There are six known D'--type (dusty) symbiotic binaries, also called
yellow symbiotics. These are binaries with F--G giants and WD
companions. Circumstellar dust and emission lines are present, and the
giants rotate rapidly.  Carbon and s--process elements are
overabundant (Pereira et~al. 2005).  Since the F--G giants in these
stars are more massive than the K giants in Ba~II stars and the
orbital periods (a few hundred days) are shorter, these systems may
also be candidate SN~Ia progenitors. Evolutionary models of these
systems need to be computed to understand if they can produce SNe~Ia
by accretion or mass transfer during Roche--lobe overflow, or during
the AGB mass--loss phase of F--G stars.

%\end{document}

\section{HYDROGEN--POOR CLOSE BINARIES}

\begin{deluxetable}{rrrrrr}
\tablecolumns{6}
\tablewidth{0pc}
\tablecaption{HYDROGEN--POOR CLOSE BINARIES}
\tablehead{

\colhead{}& \colhead{Period (days)} &\colhead{F(m)} &\colhead{$T_\mathrm{eff}$} 
&\colhead{M1$_{\odot}$} &\colhead{M2$_{\odot}$}}
\startdata
Upsilon Sgr & 137.95  & 1.44 & 11,800~K & 1.0 & 4 \\
HD 30353    &   362.8 & 3.6 & 12,500~K & 1.0 & 5 \\
LSS 4300    & 52.09  & 0.79 & 12,000~K & 1.0 & 3--4 \\
CPD -58 2721&    43 & 1 & 12,000~K    & 1.0   & 2--3 \\
\enddata
\end{deluxetable}

Only four hydrogen--poor close binariess (Upsilon Sgr; HD~30353 = KS
Per; LSS~4300; and CPD~-58~2721 = LSS~1922; see Table~3) are known,
which indicates that these are very rare systems.  In these systems
the primary is a very hydrogen--poor A supergiant and the secondary is
a hot evolved star or a MS star. These four systems are single--lined
spectroscopic binaries. The IUE UV spectra of Upsilon Sgr and HD~30353
show P--Cygni stellar wind profiles of lines of N~V, C~IV, Si~IV,
etc., indicating mass loss. The presence of N~V and C~IV lines and the
UV continuum indicates that the secondary components are hot
stars. However in the optical spectra the secondary components are not
detected. The secondary components may be obscured, because the IRAS
data indicates the presence of warm and cold circumstellar dust.  The
presence of warm--dust envelopes indicates recent mass--transfer and
mass--loss processes.  The masses of the primary and secondary
components are not well determined. High resolution, high
signal--to--noise ratio UV spectra may reveal the spectral lines of
the hot secondary components and allow the determination of
radial--velocity curves and masses. Accurate determination of the
masses is needed to compute evolutionary models of these systems.  The
mass--function [F(m)] values derived from the single--lined
spectroscopic orbits suggest that the hydrogen--poor primaries may
have CO core masses of about 1.0 to 1.3~M$_{\odot}$ with thin extended
outer envelopes. They may be in the post--AGB stage.  In Table~3 the
primary masses are all taken to be 1.0~M$_{\odot}$. Mass transfer from
the secondary components during the Roche--lobe overflow and
common--envelope phase may result in SNe~Ia (Morrison 1988; Uomoto
1986).

\section{BINARY BLUE METAL--POOR STARS AND BLUE STRAGGLERS}

\subsection{Binary Blue Metal--Poor Stars}

Among the blue metal--poor stars (spectral types A to F) several
binaries have been found. These have overabundances of carbon, lead,
and other s--process elements (Preston \& Sneden 2000, 2005; Sneden,
Preston, \& Cowan 2003). The binary blue metal--poor star CS~29497-030
has an extreme enhancement of lead, [Pb/Fe] = +3.7, the highest seen
in any star so far. The overabundances seem to have been acquired
during the AGB and post--AGB evolution of the companion stars, which
are now WDs (Preston \& Sneden 2005).  Model evolutionary calculations
are needed to see whether future Roche--lobe overflow or AGB phases of
the blue metal--poor stars may produce SNe~Ia.  Of 62 blue metal--poor
stars investigated by Preston \& Sneden (2000), two thirds are in
single--lined spectroscopic binaries with orbital periods of two to
4000 days.  Studies of large samples of blue metal--poor stars may
reveal short--period systems with total mass exceeding the
CL.

%\end{document}

\subsection{Binary Blue Straggers}

Blue stragglers are present in globular clusters, in old and young
open clusters, and in the field. Their evolution cannot be explained
by canonical stellar evolutionary models. The explanations for their
formation involve mass transfer in and/or mergers of binaries, or
stellar collisions.  All these mechanisms may be at work in globular
clusters.  Recent studies of blue stragglers in globular clusters and
in the field indicate that most of the blue stragglers are binaries
and a large fraction of them may contain WD companions.  Carney
et~al. (2005) studied the radial--velocity variations of several
metal--poor field blue stragglers, all known to be deficient in
lithium. They found all of them to be single--lined spectroscopic
binaries with periods ranging from 302 to 840 days, similar to
findings of Preston \& Sneden (2000) for other blue--straggler
candidates.  Preston \& Sneden (2000) and Sneden, Preston, \& Cowan
(2003) concluded that field blue stragglers are created almost solely
by mass transfer.  Carney et~al. (2005) argued that the secondaries in
all these systems are WDs.  They found a steeper mass function for blue
straggler binaries than for lower--mass single--lined spectroscopic
binaries, indicating a narrower range in secondary masses. They also
found that the orbital elements of all metal--poor binary blue
stragglers are consistent with stable mass transfer, which indicates
that stable mass transfer and accretion onto the WD may take place
during the Roche--lobe overflow or AGB phases of the blue
stragglers. Some of these systems with short orbital period and total
mass exceeding the CL may produce SNe~Ia.

%\end{document}

\subsection{Carbon-- and s--Process--Rich Very Metal--Poor Binaries}

From an analysis of observed radial--velocity variations of very
metal--poor and carbon--rich stars with overabundances of s--process
elements Lucatello et~al. (2005) found that the binary fraction among
these stars is higher than that found in the field, suggesting in fact
all of these objects are binaries, with WD components.  The
overabundance of carbon and s--process elements is the result of
accretion and mass transfer during the AGB stage of the present WD
companions. The orbital periods of some of these systems are found to
be short.  Long--term radial--velocity monitoring of a large sample of
these stars can lead to the accurate determination of orbital elements
and masses of the components.  Some systems with short orbital periods
and massive WDs and total mass exceeding the CL are expected. The HK
survey (Beers et~al. 1992; Beers 1999) and the Hamburg--ESO survey
(Christlieb et~al. 2001a,b; Christlieb 2003) of large samples of very
metal-poor stars have shown that up to 25 percent of stars with
metallicities lower than [Fe/H] = -2.5 are carbon--enhanced, ([C/Fe]
$>> 1.0$).  Until recently the origin of these stars and the origin of
the overabundances of carbon and s--process elements remained unclear
(Norris et~al. 1997; Aoki et~al. 2001, 2002a,b,c).  Such metal--poor
stars in binaries with short orbital periods and massive WD companions
may produce SNe~Ia.  Several of these stars have $T_\mathrm{eff}$ in
the range of 6000~K to 7000~K and surface gravities (log~g) in the
range of 3 to 4. Further radial--velocity studies and UV spectra may
enable us to estimate the masses of the WD companions. If the WD mass
is 0.6~M$_{\odot}$, then some of these systems seem to have total mass
exceeding the CL. The above described three groups of binaries with WD
companions (binary blue metal-poor stars, blue stragglers, and carbon
and s-process--rich metal--poor binaries) seem to be related.

%\end{document}

\section{AGB STARS WITH WHITE--DWARF COMPANIONS}

Mass transfer from carbon--rich AGB stars with WD components can occur
through stellar wind or Roche--lobe overflow (Iben \& Tutukov
1985). Roche--lobe overflow seems unavoidable in systems with circular
orbits and periods shorter than a few hundred days. Roche--lobe
overflow from an AGB star with a deep convective envelope leads to a
CE phase and a short orbital period (Iben 1995).
Detailed evolutionary model calculations are needed to
understand if AGB binaries with WD components can result in SNe~Ia.
An AGB star in the progenitor system of the strongly
circumstellar--interacting Type~Ia SN~2002ic was suggested by Hamuy
et~al. (2003).
              
Carbon symbiotics (carbon stars with WDs) are present in the LMC and
SMC. Further study of these systems is needed.

%\end{document}

\section{POST--AGB BINARIES}

A few post--AGB binaries are known. Very metal--poor post--AGB A and F
supergiants (Table~4) are single--lined spectroscopic binaries with
orbital periods of several hundred days (Van Winckel et~al. 1995). The
post--AGB stars in these systems have WD--like CO cores, and thin
extended envelopes of 0.01 or 0.02~M$_\odot$, and circumstellar dust
disks (Cohen et~al. 2004; Van Winckel et~al. 2000). There are a few
more post--AGB binaries with dusty circumstellar disks. In all these
systems the secondaries seem to be obscured by dusty disks.  For
the progenitor of the SN~2002ic, Kotak et~al. (2004) favor a
single--degenerate system in which the donor is a post--AGB star.

\begin{deluxetable}{rrrr}
\tablecolumns{4}
\tablewidth{0pc}
\tablecaption{POST--AGB BINARIES}
\tablehead{
\colhead{}&\colhead{Period (days)} & \colhead{e} & \colhead{f(m) }}
\startdata
HR 4049 & 429 & 0.31 & 0.143 \\ 
HD 44179 & 318 & 0.38 & 0.049 \\
HD 52961 & 1305 & 0.3 & 0.46 \\
HD 46703 & 600 & 0.34 & 0.28 \\
BD +39 4926 & 775 & \nodata & \nodata \\
HD 213985 & 259 & \nodata & 0.97 \\
\enddata
\end{deluxetable}

The bipolar protoplanetary nebula M1--92 contains a binary consisting
of a WD and a F supergiant (Arrieta et~al. 2005); an accretion disk
around the WD is suggested.  Far IR observations indicate the presence
of a large dusty torus. IRAS 08544--4431 also is a post--AGB star in a
binary with a dusty disk (Maas et~al. 2003), HD~172481 is a post--AGB
binary with a red AGB star (Reyniers \& Van Winckel 2001), and
Hen~3--1312 is a post--AGB binary (Pereira 2004).

Gordon et~al. (1998) found that the eclipsing double--lined
spectroscopic binary HD~197770 is an evolved system with at least one
of the components in the post--AGB stage. The orbital period is 99.69
days and the masses of the components are 2.9 and 1.9~M$_{\odot}$.
The spectral types of both stars are B1~V--III or near B2~III, so both
components are undermassive by about a factor of five, and thus are
evolved stars. Additional evidence of the evolved nature of HD~197770
is found in the 25, 60, 100 micron IRAS images, which show two
associated dust shells.  Chemical composition analysis of both
components from the analysis of high resolution spectra is needed. The
C, N, and O abundances may enable us to further understand the
evolutionary stage (and mass--transfer and mass--loss) of both
components. Close--binary evolutionary model calculations of this
system may enable us to understand if it will produce a SN~Ia.

%\end{document}

\section{BINARY CENTRAL STARS OF PLANETARY NEBULAE}

%\begin{deluxetable}{rrrrrrrr}
%\tablecolumns{8}
%\tablewidth{0pc}
%\tablecaption{CLOSE BINARY CENTRAL STARS OF PLANETARY NEBULAE}
%\tablehead{
%\colhead{PN} & \colhead{central star} & \colhead{Period (days) }}        
%\startdata
%Abell 41& MT Ser & 0.113 \\
%DS 1 & KV Vel & 0.357 \\
%Hf 2-2 & (MACHO var) & 0.399 \\
%Abell 63 &  UU Sge  & 0.465 \\
%Abell 46 & V 477 Lyr & 0.472 \\
%HFG 1 & V 664 Cas  & 0.582 \\
%K 1-2 & VW Pyx  & 0.676 \\
%Abell 65& \nodata & 1.0  \\
%HaTr 4  & \nodata & 1.74 \\
%Tweedy 1& BE UMa & 2.29 \\
%SuWt 2 &\nodata & 2.45 \\
%Abell 35 &BD -22 3467 & \nodata \\
%LoTr 1 &\nodata & \nodata \\
%LoTr 5 &HD 112313 & \nodata \\
%\enddata
%\end{deluxetable}
%

\begin{table}[t]
\begin{center}
\begin{tabular}{rrrrrrrr}
\multicolumn{8}{c}{Table 5. CLOSE BINARY CENTRAL STARS OF PLANETARY NEBULAE}\\
\relax\\[-1.5ex]
\tableline\tableline
\relax\\[-1.5ex]
{PN} & {central star} & { Period (days) }        \\
\relax\\[-1.5ex]
\tableline
\relax\\[-1.5ex]
Abell 41& MT Ser & 0.113 \\
DS 1 & KV Vel & 0.357 \\
Hf 2-2 & (MACHO var) & 0.399 \\
Abell 63 &  UU Sge  & 0.465 \\
Abell 46 & V 477 Lyr & 0.472 \\
HFG 1 & V 664 Cas  & 0.582 \\
K 1-2 & VW Pyx  & 0.676 \\
Abell 65& \nodata & 1.0  \\
HaTr 4  & \nodata & 1.74 \\
Tweedy 1& BE UMa & 2.29 \\
SuWt 2 &\nodata & 2.45 \\
Abell 35 &BD -22 3467 & \nodata \\
LoTr 1 &\nodata & \nodata \\
LoTr 5 &HD 112313 & \nodata \\
\relax\\[-1.5ex]
\tableline
\end{tabular}
\end{center}
\end{table}

Sixteen planetary nebulae (PNe) are known to contain close--binary
nuclei (Bond 2000) with orbital periods from 0.1 to 3~days (Table
5). The central stars (CSPNe) have CO core masses of about
0.6~M$_{\odot}$. It is estimated that 10 percent of PN nuclei may be
very close binaries (Bond \& Livio 1990). The short orbital periods
indicate that the PNe must have been ejected during CE phases.

Three PNe with binary nuclei (Abell~35, LoTr~1 and LoTr~5) have
optical spectra dominated by late G--K stars, but whose UV spectra
indicate the presence of extremely hot (100,000~K) WD companions (Bond
\& Livio 1990). The G--K stars are rapidly rotating and have active
chromospheres.  The hot WDs in these three systems have masses more
than 0.6~M$_{\odot}$. Mass loss and mass transfer from the G--K
components during the Roche--lobe overflow or AGB phases may result in
SNe~Ia.  Evolutionary model calculations need to be carried out.

Bond et~al. (2002) found that PN SuWt~2 is a triple system consisting
of two A stars with 4.9~day orbital period and a hot CSPN.  The PN
appears as an elliptical ring with much fainter bipolar lobes. Bond
et~al. (2002) suggest that SuWt~2 and V~471~Tau (\S~2.1) are exotic
descendants of triple systems. The masses of the A stars in SuWt~2 are
both near 2~M$_{\odot}$, so the initial mass of the CSPN was more than
2~M$_{\odot}$.

G135.6+01.0 = WeBo~1 contains a close binary consisting of a
late--type giant and a hot CSPN (Bond et~al. 2003). The giant is
overabundant in carbon and s--process elements which indicates that
mass transfer and/or accretion has taken place during the AGB stage of
the CSPN.  Further study of WeBo~1 and SuWt~2 including the
measurement of orbital periods and masses of the components may enable
us to understand if such close--binary CSPNe can produce SNe~Ia.
        
Wolf--Rayet (WC) CSPNe are hydrogen deficient and overabundant in
helium and carbon. About 10 to 15 percent of CSPNe are of type
WC. From ISO observations, a dual dust chemistry (oxygen-- and
carbon--rich) is almost exclusively associated with WC
CSPNe. Oxygen--rich dust resides in a disk, while the carbon-rich dust
is more widely distributed. HST STIS spectroscopy of the WC CSPN
CPD~-56~8032 indicates the presence of a dust disk or torus. All the
WC PNe are IRAS sources. From a radial--velocity survey of 18 WC CSPNe
De~Marco et~al.(2004) found 8 WC stars showing radial--velocity
variations, indicating binarity. De~Marco et al. concluded that all WC
CSPNe are binaries. Evolutionary model calculations are needed to
understand if binary WC CSPNe can produce SNe~Ia.  

G135.9+55.9, a PN in the Galactic halo, contains a close--binary
nucleus with short orbital period (Tovmassian et~al. 2004).  From
radial--velocity data the mass of the WD is about 0.9~M$_{\odot}$ and
the mass of the CSPN is about 0.55~M$_{\odot}$; the total mass exceeds
the CL and this system may produce a SN~Ia.

%\end{document}

\section{SYMBIOTIC STARS}

Symbiotic stars are interacting binaries consisting of a red giant and
a WD. The D--type symbiotics have warm circumstellar dust and the cool
components of many of these are Mira variables. Some of these have
bipolar nebulae. There are several symbiotics in which the donor
stars are of types G--K. The total mass of such systems exceeds the
CL. The spectra reveal accretion and mass transfer
from the G--K companion to the WD.  We need to derive accurate orbital
periods, masses, and accretion and mass transfer rates. Since the
G--K--M stars in the symbiotics are AGB stars the orbital periods are
longer. In several of the D--type symbiotics the pulsating Mira
variables show evidence for mass loss and for the presence of cool and
warm circumstellar dust. The WDs in D--type symbiotics with periods in
the range of 200 to 600 days may have relatively high accretion rates,
as the mass--loss rates from the companion Miras are on the order of
10$^{-7}$~M$_{\odot}$~y$^{-1}$.

Mira is an interacting binary consisting of a Mira--variable AGB star
(Mira~A) and an accreting WD (Mira~B). Mira~A is losing mass at a high
rate ($\sim10^{-7}$~M$_{\odot}$~y$^{-1}$).  The system has been
studied using HST and CHANDRA (Karovska et~al. 2005). The symbiotic
binaries CH~Cyg and R~Aqr are similar to Mira. Evolutionary model
calculations of these systems are needed to understand if they will
produce SNe~Ia. The large X--ray outburst in Mira~A detected by
CHANDRA may have been caused by a magnetic flare followed by a large
mass ejection (Karovska et~al. 2005). Such large--scale mass ejections
from Mira~A may result in an increased accretion rate onto the WD.  If
they occur often, such events in Mira and in other symbiotics may
enable the WDs to reach the CL.  Munari \& Renzini
(1992) argue that only 4~percent of the symbiotics in the Galaxy need
to evolve to SNe~Ia in order to account for the observed SN~Ia rate.
However Kenyon et~al. (1993) countered that symbiotics probably cannot
produce SNe~Ia at a significant rate (see also Yungelson et~al. 1995).

Some symbiotics (R~Aqr, AG~Dra, RR~Tel, CD-43~14304, and J0048 and
Ln358 in the SMC) are supersoft X--ray sources (SSXSs; Greiner
1996). There seems to be two groups of symbiotic SSXSs: symbiotic
novae (e.g., RR~Tel, J0048) and quiescent symbiotics (AG~Dra and most
others). Symbiotic novae have strong soft X--ray emission during their
optical outburst.  On the contrary, AG~Dra displays soft X--ray
emission in its quiescent state (Greiner et~al. 1997). It is not clear
if the WDs in some of these systems can gain enough mass to reach the
CL. In the LMC and SMC the WDs in some of the symbiotics have high
luminosities (3000~L$_{\odot}$) and effective temperatures (130,000~K)
which indicates that they are at the upper end of the luminosity
function (M\"urset et~al. 1996).  There are about six symbiotics in
the SMC with WD temperatures in the range 1.3 to $2 \times 10^{5}$~K,
WD masses of about 0.7~M$_{\odot}$ (Iben 1982), and red--giant masses
of about 3~M$_{\odot}$.  Kahabka (2004) studied the six--year
supersoft X-ray light curve of the symbiotic nova SMC~3 =
RXJ0048.4-7332. He found that the 0.8~M$_{\odot}$ WD is undergoing
steady nuclear burning with an accretion rate close to
10$^{-7}$~M$_{\odot}$~y$^{-1}$, and he suggested that it may become a
sub--CL SN~Ia (although it is doubtful that sub--CL WDs can
explode). Monitoring of the following symbiotic binaries may enable us
to understand if some of them can produce SNe~Ia: Mira, CH~Cyg,
V741~Per, Wray~157, AS201, BD--21~3873, He2-467, S190, and the
symbiotic SSXSs RR~Tel, AG~Dra, R~Aqr, CD~-43~14304, and the SMC
symbiotic SSXSs J0048, Ln358, and SMC 3 = RXJ0048.4-7332.

%\end{document}

\section{DWARF NOVAE}
   
Cataclysmic variables (CVs) with accreting WDs consist of several
types, such as classical novae, recurrent novae, nova--like variables,
dwarf novae, helium~CVs, and magnetic~CVs. The total mass in most CV
systems is much lower than the CL. There are a few dwarf novae in
which the WD is about 1~M$_\odot$: GK~Per, SS~Aur, HL~CMa, U~Gem,
Z~Cam, SY~Cnc, OY~Car, Z~Cam, TW~Vir, AM~Her, SS~Cyg, RU~Peg, GD~552,
and IP~Peg.  The secondaries are K and M stars. A few of these systems
with early K~type secondaries may have total mass close to the
CL. Accurate determination of masses of the components in these
systems is needed. King et~al. (2003) considered the CVs with massive
WDs as candidate SN~Ia progenitors.  From multiwavelength studies of
the properties of the accreting WDs in dwarf novae during quiescence
we may be able to estimate the WD masses.  RU~Peg has a massive WD
(1.29~M$_{\odot}$) and the secondary is a K2~V star. The orbital
period is 8.99 hours (Stover 1981; Wade 1982; Shafter 1983). The
near--Chandrasekhar mass of the WD has been corroborated by the sodium
8190 \AA\ radial--velocity study of RU~Peg by Friend et~al. (1990),
who obtained a WD mass of 1.24~M$_{\odot}$ and also found very good
agreement with the solution, including the range of plausible
inclinations, of Stover (1981). The total mass of the system exceeds
the CL. RU~Peg undergoes dwarf nova outbursts every 75 to 85 days,
with outburst amplitudes of three magnitudes and durations of three
days (Ritter \& Kolb 1998). Between outbursts RU~Peg has a steeply
rising far--UV continuum.  From a detailed analysis of IUE spectra of
RU~Peg during quiescence Sion \& Urban (2002) found the WD
$T_\mathrm{eff}$ to be 50,000~K. They conclude that RU~Peg contains
the hottest WD yet found in a dwarf nova. The WD of RU Peg is
15,000~K, hotter than the hottest WDs in dwarf novae (U~Gem and
RX~And) above the period gap. RU~Peg's WD is in the temperature regime
occupied by the WDs in the nova--like variables UX~UMa and VY~Scl. The
accretion disk in RU~Peg may be among the largest of any dwarf nova
(Sion \& Urban 2002). The accreting WD in RU~Peg with $T_\mathrm{eff}$
= 50,000~K should have an envelope thermal structure that could
support thermonuclear burning. Further studies of RU~Peg and similar
systems with massive WDs are needed to understand if they evolve to
SNe~Ia. There may be short and long term variations in the accretion
rates, i.e., at times it may be only 1 to 2$\times 10^{-9}$
M$_{\odot}$~y$^{-1}$ and at other times as high as 10$^{-7}$
M$_{\odot}$~y$^{-1}$. Further study of RU~Peg--type systems with
massive WDs may enable us to understand if some of them can become
SNe~Ia.

From SDSS data Szkody et~al. (2006) discovered a large number of new
CVs, including many interesting systems such as eclipsing, pulsating,
and magnetic CVs.  This database is a good resource for population and
evolution studies of CVs, pre--CVs, and WD companions.  Detailed
studies of these systems are needed.

%\end{document}

%\begin{deluxetable}{rrr}
%\tablecolumns{3}
%\tablewidth{0pc}
%\tablecaption{RECURRENT NOVAE WITH MASSIVE WHITE DWARFS}
%\tablehead{
%\colhead{}&\colhead{ WD mass (M$_{\odot}$)}&\colhead{Period (days)}}
%\startdata
%T CrB & 1.37 & 227.67\\
%RS Oph & 1.35 & 460\\
%V 745 Sco & 1.35 & \nodata \\
%V3890 Sgr & 1.35 & \nodata \\
%U Sco & 1.37 & 1.23056 \\
%V 394 CrA & 1.37 & 0.7577 \\
%IM Nor & \nodata & \nodata \\
%CI Aql & \nodata & \nodata \\
%LMC 1990-2 & \nodata & \nodata \\
%\enddata
%\end{deluxetable}

\section{RECURRENT NOVAE}

Recurrent novae, in which a WD accretes from a red giant, have been
considered as candidate SN~Ia progenitors (Starrfield, Sparks \&
Truran 1985; Hachisu \& Kato 2001). From detailed modeling of the
decline of the outburst light curves of T~CrB, RS~Oph, V745~Sco, and
V3890~Sgr, Hachisu \& Kato (2001) suggest that the WDs are approaching
the CL (Table 6) and will become SNe~Ia.  Wood--Vasey \& Sokolowski
(2006) proposed that a shell ejected by a recurrent nova created the
evacuated region surrounding SN~2002ic.  They suggest that recurrent
novae are SN~Ia progenitors, with the periodic shell ejections
creating structure in the circumstellar region.

RS~Oph is one of the well--observed recurrent novae and is suggested
to be a SN~Ia progenitor system (Hachisu \& Kato 2001).  It underwent
its sixth recorded outburst on 2006 February~12 (Narumi
et~al. 2006). Detailed multi-wavelength study of the the recent
outburst of RS~Oph (Hachisu et~al. 2006; Monnier et~al. 2006;
Sokoloski et~al. 2006) and models (Hachisu, Kato, \& Luna 2007)
suggest a WD of $1.35 \pm 0.01$~M$_{\odot}$, and that the WD is
growing in mass at an average rate of about $1 \times
10^7$~M$_{\odot}$~yr$^{-1}$, which indicates that RS~Oph will produce
a SN~Ia in a few hundred thousand years. Recurrent novae U~Sco and
CI~Aql appear to be similar to RS~Oph.

\begin{table}[t]
\begin{center}
\begin{tabular}{rrr}
\multicolumn{3}{c}{Table 6. RECURRENT NOVAE WITH MASSIVE WHITE DWARFS}\\
\relax\\[-1.5ex]
\tableline\tableline
\relax\\[-1.5ex]
{}&{ WD mass (M$_{\odot}$)}&{Period (days)}\ \ \ \\
\relax\\[-1.5ex]
\tableline
\relax\\[-1.5ex]
T CrB & 1.37 & 227.67\\
RS Oph & 1.35 & 460\\
V 745 Sco & 1.35 & \nodata \\
V3890 Sgr & 1.35 & \nodata \\
U Sco & 1.37 & 1.23056 \\
V 394 CrA & 1.37 & 0.7577 \\
IM Nor & \nodata & \nodata \\
CI Aql & \nodata & \nodata \\
LMC 1990-2 & \nodata & \nodata \\
\relax\\[-1.5ex]
\tableline
\end{tabular}
\end{center}
\end{table}

\section{VY SCULPTORIS NOVA--LIKE SYSTEMS}

Long--term monitoring of the transient SSXS RXJ0513.9-6951 (Schaeidt
et~al. 1993) has revealed quasi-periodic optical intensity dips of
about four weeks duration (Reinsch et~al. 1996; Southwell
et~al. 1996), which occur during the X-ray on states. The similarity
of the optical behaviour of this SSXS with that of VY~Scl stars led
Greiner et~al. (1999) to suggest that some of the VY~Scl stars may
indeed be hitherto unrecognized SSXSs. They observed V~751~Cyg with
ROSAT HRI and found transient supersoft X--ray emission and
anti--correlation of X--ray and optical intensity. The X--ray emission
is very luminous and the spectrum is very soft, similar to spectra of
SSXSs such as RXJ0513.9-6951.

From the Balmer--line radial--velocity curve of VY~Scl, Martinez-Pais
et~al. (2000) found the orbital period to be 0.232 days. They also
found evidence for the presence of a third component. They estimate
the mass of the WD to be 1.22~M$_{\odot}$ and the secondary star mass
to be 0.43 M$_{\odot}$. The orbital period of the third star is about
5.8 days. The mass and evolutionary status of the third component is
not clear; it may be a normal star of about 0.8~M$_{\odot}$ or a
low--luminosity higher mass star, in which case it is probably a
compact object (Martinez--Pais et~al. 2000).  The total mass of the
0.232--day orbital period binary is 1.65~M$_{\odot}$ WD so it may
become a SN~Ia.

The detection of supersoft X--ray emission from the VY~Scl--type star
V~751~Cyg (Greiner et~al. 1999) and the detection of a massive WD in
VY~Scl with 0.232~day orbital period (Martinez--Pais et~al. 2000)
indicate that some of the VY Scl type systems may become
SNe~Ia. Greiner (1998) studied the ROSAT observations of VY Scl stars
to investigate the presence, strength and spectrum of soft X-ray
emission (0.1 - 2.4 keV). He found soft X-ray emission from 9 out of
the 14 VY Scl stars that are detected with ROSAT. Further optical
studies of these nova-like stars and determination of their orbital
periods and WD masses, and further studies in the X-ray region may
enable us to detect some of the systems that may become SNe~Ia.

%\begin{deluxetable}{rrrrr}
%\tablecolumns{5}
%\tablewidth{0pc}
%\tablecaption{SUPERSOFT X--RAY BINARIES}
%\tablehead{
%\colhead{Object} & \colhead{Period (days)} & \colhead{ V(max)}&\colhead{ V(min)}
%&\colhead{ L(WD ergs/sec)}}
%\startdata 
%Galaxy &        &      &       &   \\
%QR And & 0.6605 & 11.5 & 12.65 & 4 $\times10^{36}$\\
%MR Vel  & 4.0288 & 17.1 & 17.3 & 4 $\times10^{35}$\\ 
%LMC     &        &      &      &   \\
%J0513    & 0.7628 & 16.4 & 17.55 & 7 $\times10^{37}$ \\
%J0439   & 0.1404  & \nodata & 21.74 & 9 $\times10^{37}$\\
%J0537    & 0.1458 & \nodata & 19.66 & \nodata \\
%CAL 83  & 1.0417  & 16.9 &  17.5 & 2 $\times10^{37}$\\
%CAL 87  & 0.4425  & \nodata &  18.9 & 5 $\times10^{36}$\\
%SMC     &         &       &      &  \\
%1E0035   & 0.1719 & \nodata   & 20.3 & 5 $\times10^{36}$\\
%\enddata
%\end{deluxetable}

%\end{document}

\section{SUPER--SOFT X--RAY SOURCES}

SSXSs (Table~7) contain accreting WDs (Van~den~Heuvel
et~al. 1992). They show soft X--ray spectra with almost all of the
flux below 0.5 to 1~keV (Tr\"umper et~al. 1991) and luminosities close
to the Eddington limit of a 1~M$_\odot$ star (see the review by
Kahabka \& Van~den~Heuvel 1997). Most of the known SSXSs are located
in the LMC, SMC, and M31. In the Galactic disk, the high interstellar
absorption limits the distance of detectable SSXSs to about
2~kpc. Most of the SSXSs have optical properties reminiscent of
low--mass X--ray binaries, which suggests that their optical and UV
spectra are strongly affected by soft X--ray irradiation from the
accreting WD companion. ROSAT, BEPPOSAX, and ASCA X--ray spectra of
most SSXSs indicate that the accreting WDs are very hot. 

\begin{table}[h]
\begin{center}
\begin{tabular}{rrrrr}
\multicolumn{5}{c}{Table 7. SUPERSOFT X--RAY SOURCES}\\
\relax\\[-1.5ex]
\tableline\tableline
\relax\\[-1.5ex]
{Object} & {Period (days)} & { V(max)}&{ V(min)}
&{ L WD (erg~s$^{-1}$)}\\
\relax\\[-1.5ex]
\tableline
\relax\\[-1.5ex]
Galaxy &        &      &       &   \\
QR And & 0.6605 & 11.5 & 12.65 & 4 $\times10^{36}$\\
MR Vel  & 4.0288 & 17.1 & 17.3 & 4 $\times10^{35}$\\ 
LMC     &        &      &      &   \\
J0513    & 0.7628 & 16.4 & 17.55 & 7 $\times10^{37}$ \\
J0439   & 0.1404  & \nodata & 21.74 & 9 $\times10^{37}$\\
J0537    & 0.1458 & \nodata & 19.66 & \nodata \\
CAL 83  & 1.0417  & 16.9 &  17.5 & 2 $\times10^{37}$\\
CAL 87  & 0.4425  & \nodata &  18.9 & 5 $\times10^{36}$\\
SMC     &         &       &      &  \\
1E0035   & 0.1719 & \nodata   & 20.3 & 5 $\times10^{36}$\\
\relax\\[-1.5ex]
\tableline
\end{tabular}
\end{center}
\end{table}

In the model proposed by Van~den~Heuvel et~al. (1992) the relatively
massive accreting WD sustains steady burning of the hydrogen--rich
accreted material from a MS or subgiant donor star. They suggested
that the accretion has to occur at a finely tuned rate, in the range
1.0 to $4.0 \times 10^{-7}$ M$_{\odot}$~y$^{-1}$. At lower rates,
hydrogen burning is unstable and occurs in flashes, while at higher
rates an extended envelope forms. Rappaport et~al. (1994) studied the
formation and evolution of such binaries, reproducing their typical
luminosities, effective temperatures and orbital periods. They found
that there should be more than 1000 SSXSs in the Galaxy and in M31 and
about 100 in the LMC with properties that closely match those of the
observed SSXSs.  They found that the orbital periods should be in the
range of 0.3 days to 1.4~days, and the WDs should have masses in the
range of 0.7 to 1.05~M$_{\odot}$ and effective temperatures in the
range of 1 to $5 \times 10^{5}$~K.  The masses of donor stars are in
the range of 1.3 to 2.7~M$_{\odot}$. Rappaport et~al. (1994) estimated
the rate of Galactic SNe~Ia resulting from the evolution of SSXSs to
be 0.006~y$^{-1}$.

Ivanova \& Taam (2004) studied the fate of close binary systems
(orbital periods 1 to 2 days) that consist of evolved MS donors (1 to
3.5~M$_{\odot}$) and WD companions (0.7 to 1.2~M$_{\odot}$) and that
undergo a phase of mass transfer on a thermal time scale, allowing for
the possibility of an optically thick wind driven from the WD.  To
evolve toward a SN~Ia they found that the WD should be relatively
massive (more than 0.8~M$_{\odot}$) and the donor star needs to be
above 2~M$_{\odot}$ (but the donor/WD mass--ratio needs to be smaller
than three). With these conditions the mass--transfer rates are
sufficiently high that surface hydrogen burning provides the bulk of
the energy, and the sources are likely to be observed as SSXSs.

Bitzaraki et~al. (2004) performed evolutionary calculations that led
to the formation of luminous SSXSs. They found that the progenitors of
the WDs had MS masses $\sim$7~M$_{\odot}$ and companions in the range
of 1.5 to 3.0~M$_{\odot}$. In their calculations they included
thermohaline mixing after mass transfer during binary evolution.  They
concentrated on early case--C evolution which means that the primary
fills its Roche lobe when it ascends the early asymptotic giant branch
while its core is highly evolved and massive enough to form a CO
WD. Their models accounted for the observed properties of SSXSs such
as U~Sco and very luminous extragalactic SSXSs such as CAL~83 in the
LMC and the CHANDRA source N1 in M81.

Lanz et~al. (2005) did a NLTE model--atmosphere analysis of the LMC
SSXS CAL~83 using X--ray spectra obtained with the CHANDRA
high--resolution camera and low--energy transmission grating, and
XMM--Newton spectra.  They found a very rich absorption--line spectrum
from the hot WD photosphere but no spectral signature of a wind. They
obtained the first direct spectroscopic evidence that the WD is
massive (1.3~M$_{\odot}$).  They found that low--mass models do not
match the observed spectrum as well as high-mass models, and they
concluded that 1.0~M$_{\odot}$ is a robust lower limit for the WD mass
in CAL~83. They found $T_\mathrm{eff}$ to be 550,000~K and log~g =
8.5. They also found that the short timescale of the X--ray off states
(about 50 days; Kahabka 1998) is consistent with a high WD mass. Their
analysis of the spectrum of CAL~83 provides direct support for SSXSs
as likely progenitors of SNe~Ia.
 
Others also have considered SSXSs as SN~Ia progenitors (Branch
et~al. 1995; Starrfield et~al. 2004; Van~den~Heuvel et~al.  1992;
Kahabka \& van~den~Heuvel 1997; Hachisu et~al. 1999). Hachisu et~al.
found that the accreting WDs will have strong and hot stellar winds
when the mass accretion to the WD exceeds a critical rate. The excess
matter transferred above the critical rate is expelled by stellar
winds. In such a situation the WD can grow to the CL resulting in a
SN~Ia. This model, in which a CE does not develop, is called accretion
wind evolution.  The SSXSs RX~J0513.9-6951 and V~Sge and a few more
SSXSs (Starrfield et~al.  2004) are suggested to be examples of SN~Ia
progenitors via accretion wind evolution.  A list of SSXSs and some of
their parameters are given in Table~7.

Some symbiotic binaries are SSXSs (e.g., AG~Dra; Greiner
et~al. 1997). A search in the ROSAT archive by M\"urst et~al. (1996)
enabled them to sort the symbiotic stars into three types based on the
hardness of their X--ray spectra. Symbiotic binaries Ln~358, Dra~C-1,
R~Aqr, AG~Dra, RR~Tel (a symbiotic nova), and CD~-43~14304 were found
to show SSXS characteristics. There are a few symbiotic binaries with
harder X--ray specta, which may be due to colliding hot winds. Some of
the symbiotic binaries with the hardest X--ray spectra may contain
massive accreting WDs or accreting neutron stars.

Some classical and recurrent novae have been detected as luminous
SSXSs during the late--decline phase of their outbursts (e.g., V~1974
Cyg = Nova Cyg 1992, Balman et~al. 1998; GQ~Mus = Nova Mus 1983,
Shanley et~al. 1995; and U~Sco, Kahabka et~al. 1999).

The recurrent nova U~Scorpii is found to be a SSXS with a hot and
massive accreting WD (Kahabka et~al. 1999). The WD $T_\mathrm{eff}$ is
estimated to be $9 \times 10^{5}$~K. The donor is a MS star of
1.5~M$_{\odot}$. The system is an eclipsing binary with an orbital
period of 1.23~days (Schaefer \& Ringwald 1995; Thoroughgood
et~al. 2001). The mass of the WD is 1.37~M$_{\odot}$ (Bitzaraki
et~al. 2004; Thoroughgood et~al. 2001), very close to the CL. U~Sco
shows nova outbursts almost every eight years. It has shortest known
recurrence period (Iijima 2002). The outburst mechanism for the U~Sco
subclass of recurrent novae is similar to the thermonuclear runaway
model of classical nova outbursts (Starrfield, Sparks \& Truran
1985). The helium abundance in the ejecta of U~Sco (Iijima 2002) is
similar to that found in the ejecta of normal classical novae. Some
authors have estimated higher helium abundance in the ejecta of U~Sco
(Barlow et~al. 1981). It is important to derive an accurate heluim
abundance from analysis of high--resolution spectra in order to find
out if the donor star is a normal star or a helium--rich
star. Thoroughgood et~al. (2001) proposed that U~Sco is the best SN~Ia
progenitor currently known and estimated that it will be a SN~Ia in
$\sim7 \times 10^{5}$ years.  From evolutionary calculations Bitzaraki
et~al. (2004) also consider U~Sco to be a good SN~Ia candidate.

%\end{document}

\section{V SAGITTAE--TYPE CLOSE BINARIES}

Most of the known SSXSs are located in external galaxies (Greiner
1996), which makes detailed optical observations difficult. It is
therefore of great interest to identify galactic SSXSs and related
stars. X--ray surveys have not been able to detect many Galactic SSXSs
because the soft X--ray emission is strongly attenuated by the
interstellar medium. It has been suggested that V~Sge--type close
binary stars (V~Sge, V617~Sgr, HD~104994, WX~Cen, and T~Pyxidis; see
Table~8) have spectroscopic and photometric properties that are very
similar to those of SSXSs (Steiner \& Diaz 1998; Patterson et
al. 1998; Greiner \& van~Teeseling 1998; Simon 2003). This suggestion
is based on characteristics which are typical for SSXSs, but are rare
or even absent among canonical CVs: (1) the presence of both O~VI and
N~V emission lines, (2) a He~II (4686\AA) to H$_{\beta}$
emission--line ratio greater than 2, (3) high luminosities and very
blue colors, and (4) orbital light curves that are characterized by
wide, deep eclipses similar to those observed in some of the SSXSs
(Hachisu 2004).

\setcounter{table}{7}
\begin{deluxetable}{rrrrrr}
\tablecolumns{6} \tablewidth{0pc} \tablecaption{V Sge--CLOSE BINARIES}
\tablehead{ \colhead{}& \colhead{Period (hours)}&\colhead{
V}&\colhead{(B-V)}&\colhead{(U-B)} &\colhead{W-R type}} \startdata V
Sge & 12.34 & 11.6 & 0.0 & -0.9 & WN5\\ V617 Sgr & 4.97 & 14.8 & -0.04
& -0.87 & WN5\\ HD 104994 & 7.46 & 10.9 & 0.06 & -0.84 & WN3(pec)\\ WX
Cen & 10.0 & 13.7 & 0.4 & -0.7 & WN7\\ HD 45166 & 0.357 & 9.88 &
\nodata & \nodata & \nodata \\ T Pyxidis & 1.92 & \nodata & \nodata &
\nodata & \nodata \\ \enddata
\end{deluxetable}

V~Sge is a very blue star that varies in brightness from 9.6 to 14.7
magnitudes with a mean around 11.6. It is an eclipsing and
double--lined spectroscopic binary with a period of 0.5142~days. The
WD mass is estimated to be 1.0 to 1.3~M$_{\odot}$ and the mass of the
companion is estimated to be 2 to 3~M$_{\odot}$. The complex nature of
the variable light curve and spectrum makes it difficult to derive
accurate parameters.  The optical spectrum shows complex emission
lines (Patterson et~al. 1998; Steiner \& Diaz 1998). Greiner \&
van~Teeseling (1998) studied the X--ray properties of V~Sge and its
relation to the SSXSs. They found that during optically bright states,
V~Sge is a faint, hard X--ray source, while during optically faint
states (V magnitude fainter than 12), it has properties similar to
those of SSXSs. They explain the different optical and X--ray states
by a variable amount of extended uneclipsed matter, which during the
optically bright states contributes significantly to the optical flux
and completely absorbs the soft X--ray component. An additional,
perhaps permanent, hard X--ray component seems to be present in order
to explain the X--ray properties during the optically bright, hard
X--ray state. The anti--correlation of soft X--ray emission with
optical brightness of V~Sge is similar to that observed in
RXJ0513.9-6951, a transient SSXS in the LMC (Reinsch et~al. 1996;
Southwell et~al. 1996).  RXJ0513.9-6951 turns on as a SSXS only during
1~magnitude optical dips, which occur every 100 to 200 days and last
about 30 days. This behaviour has been explained by assuming expansion
and contraction of the shell--burning WD. The model that has been
suggested for RXJ0513.9-6951 cannot explain the observed X--ray data
of V~Sge: the optical brightness changes of V~Sge are very rapid ---
both the faint--bright--state transitions as well as the succession of
different faint states may occur on timescales of less than 1 day.

Hachisu (2004) proposed a model to explain the long--term light--curve
variations of the SSXSs V~Sge and RXJ0513.9-6951 based on an optically
thick wind model of mass--accreting WDs. The observed long--term light
curves of V~Sge and RXJ0513.9-6951 are very similar (Hachisu
2004). When the mass accretion rate exceeds the critical rate of
$1\times 10^{-6}$~M$_{\odot}$~y$^{-1}$, optically thick strong winds
begin to blow from the WD, and the WD can accrete and burn
hydrogen--rich matter on the surface at the critical rate.  The excess
matter is expelled in the wind and the WD can grow to the CL resulting
in a SN~Ia. Using the accretion wind evolution model they were able to
reproduce the transition between the optically high (X--ray--off)
state and the optically low (X--ray--on) states of RXJ0513.9-6951 and
V~Sge.

WX~Cen = WR~48c is a V~Sge type close binary with an orbital period of
0.417 days. The light curve and spectrum of WX~Cen are similar to
those of V~Sge. Based on the accretion rate
(10$^{-7}$~M$_{\odot}$~y$^{-1}$), the WD mass (1.16~M$_{\odot}$) and
the short orbital period Oliveira \& Steiner (2004) estimated that in
$\sim5 \times 10^{6}$~years WX~Cen may become a SN~Ia..

Two other V~Sge--type close binaries are V~617~Sgr and HD104994 =
DI~Cru. V 617~Sgr is a close binary with an orbital period of 4.97
hours. The light curve and spectrum of V~617 Sgr are very similar to
V~Sge. X--ray emission has been detected from HD~104994. The orbital
period is 6.78 hours. The light curve, its asymmetries, and the
flickering of HD~104994 are similar to those of V~Sge and V~617 Sgr.
Steiner et~al. (2006) find that the orbital period of V~617 Sgr to
evolve quite rapidly which is consistent with the idea that V~617 Sgr
is a wind driven accretion supersoft source. They consider that it is
similar to the wind driven supersoft X-ray binaries SMC 13 and T
Pyx. They consider that V~617 Sgr will evolve into a Type Ia supernova
in few million years.  All the above mentioned V~Sge--type systems
have close similarities to SSXSs.  Patterson et~al. (1998) considered
T~Pyxidis to be similar to V~Sge--type stars and SSXSs.  T~Pyx has a
high--excitation spectrum, very blue colors, and very high luminosity
similar to that of V~Sge (Patterson et~al. 1998).

Further multiwavelength long--term spectroscopic and photometric
monitoring of these systems is needed to derive accurate masses of the
WDs and donor stars, and accretion and mass--loss rates.  Also, from
an analysis of SDSS spectra we may be able to detect new V~Sge--type
systems with optical spectra similar to those of V~Sge and SSXSs. A
few hundred V~Sge--type systems are expected to exist in the
Galaxy. Since the optical spectra show O~VI, N~V, and He~II it may be
possible to detect new V~Sge--type systems in SDSS data using the
V~Sge spectrum as a template.

%\end{document}

\section{BINARITY OF WHITE DWARFS WITH HARD X--RAY EMISSION}
    
WDs themselves do not emit hard (greater than 0.5 keV)~X--rays but if
a WD accretes material the released gravitational energy may power
hard X--ray emission.  The most likely source to provide material for
accretion onto a WD is a binary companion. Alternatively, if a WD has
a binary companion with strong coronal activity, the hard X--ray
emission may come from the companion. Either way, hard X--ray emission
from WDs implies the presence of binarity.  Fleming et~al. (1996)
found nine DA WDs with hard X--ray emission using the ROSAT all--sky
survey, and all nine have late--type (F to M) companions. O'Dwyer
et~al. (2003) made a systematic search for hard X--ray emission from
WDs by correlating the WD catalog of McCook \& Sion (1999) and the
ROSAT point--source catalog of White et~al. (2000).  They found 76 WDs
coincident with X--ray sources at a high level of
confidence. (Multiwavelength studies and radial--velocity monitoring,
and near-IR photometry of WDs with hard X--ray emission may reveal
late--type companions).  Among these sources, 17 show significant hard
X--ray emission at energies greater than 0.5~keV, and 12 are in known
binaries; in two of these the accretion of the close companion's
material onto the WD produces the hard X--rays, and in the other 10
the late--type companion's coronal activity plus accretion of material
from coronal ejections and stellar winds seem to cause the hard
X--rays. Some of the WDs with hard X--ray emission are found to be the
hottest in the sample. Chu et~al. (2004) used an updated list of WDs
and the final ROSAT point--source catalog to find 47 new X--ray
sources convincingly coincident with WDs. Five of these have hard
X--ray emission and late-type companions.  From further
multiwavelength studies of WDs with hard X--ray emission we may be
able to find some accreting systems with F, G, and K companions.

%\end{document}

\section{DOUBLE DEGENERATES}               

Several systematic searches for double WD (DD) binaries have been made
(Napiwotzki et~al. 2004). The radial--velocity surveys have found
about 120 DDs. They found only one massive DD system, with total mass
(1.24~M$_{\odot}$), still about 10 percent below the CL. Also the long
orbital period of 12.5~hours shows that it is not a good candidate for
SN Ia within a Hubble time.  None of the systems qualify as SN~Ia
progenitors because the total mass is much smaller than the CL. In
fact, most of the individual masses seem to be smaller than
0.45~M$_\odot$, the approximate core--mass limit for helium ignition,
therefore they are helium WDs.

The only likely SN~Ia progenitor in this sample is not a DD, but the
sdB~+~WD binary KPD~1930+2752 (Maxted et~al. 2000). The orbital period
is 2.283 hours, the mass of the sdB star is 0.55~M$_{\odot}$, and the
mass of the WD is 0.97~M$_{\odot}$. The system may merge in less than
0.2 Gyr, perhaps resulting in a SN~Ia. Geier et~al. (2006) analyzed
time resolved spectroscopy of KPD 1930+2752 and found that the total
mass exceeds the CL. The total mass and the merging time of the binary
indicate that it is a very good candidate for a SN Ia progenitor
(Geier et~al. 2006).

\section{EVOLUTIONARY MODELS}

Langer et~al. (2000) studied the evolution of CO~WD and MS (single
degenerate; SD) systems that may become SNe~Ia. They found that WDs
with initial mass as small as 0.7~M$_{\odot}$ can produce SNe~Ia. They
found an upper limit for donor stars of 2.3~M$_{\odot}$. Langer
et~al. (2000) limited the maximum possible wind mass--loss rate to
three times the Eddington limit of the accreting WDs. They considered
only case~A mass transfer.

Recently Han \& Podsiadlowski (2004) also carried out a detailed study
of evolutionary models of CO~WD and MS systems, using Eggleton's
stellar evolution code. They performed binary stellar evolution
calculations for about 2300 close WD binaries and mapped the initial
parameters in the orbital period - secondary mass plane, that may
become SNe~Ia. They confirm the result of Langer et~al. that WDs with
mass as low as 0.67~M$_{\odot}$ can accrete efficiently and reach the
CL. They did not limit the mass--loss rate and they considered both
case~A and case~B mass transfer. Their upper limit for donor stars was
3.7~M$_{\odot}$.

Belczynski et~al. (2005) used the StarTrack population synthesis code
to discuss potential progenitors of SNe~Ia. They found that the SD
scenario can explain the observed delay times for models with low
common--envelope efficiency. H\"oflich et~al. (1996) and Nugent
et~al. (1997) found that the observed properties of SNe~Ia favor WDs
at the CL, while the sub--CL mass models do not explain even
subluminous, red SN Ia.

Fedorova, Tutukov \& Yungelson (2004) considered scenarios for the
evolution of close binaries resulting in the formation of
semi--detached systems in which a WD can reach the CL by accretion
from a MS or SG companion of mass~2 M$_{\odot}$. From population
synthesis studies of these systems, they found that the model occurrence
rate of SNe~Ia in semi-detached systems is $0.2 \times
10^{-3}$~y$^{-1}$, less than 10 percent of the observational estimate
of the Galactic occurrence rate of SNe~Ia.  Thus, in their model, this
channel for formation of progenitors of potential SN~Ia is not able to
produce more than 10 percent of all SNe~Ia in our Galaxy.

Greggio (2005) presented formalism to relate the rate of SNe~Ia in
stellar systems to their star--formation history through two
fundamental characteristics of the SN~Ia progenitor model : the
realization probability of the SN~Ia scenario from a single--age
stellar population, and the distribution function of the delay times,
which is proportional to the SN~Ia rate after an instantaneous burst
of star formation. Greggio (2005) suggests that different channels
(SD~CL, sub--CL, and DD~CL) could contribute to SNe~Ia, each with its
own probability, as in the realizations of the population synthesis
models. Some of the differences in the observed properties of SNe~Ia
seem to support the above mentioned notion (Benetti et~al. 2005). The
different luminosities at maximum, the different light--curve decline
rates, and the differences in parent galaxies may be due to different
typical progenitors. If both SD and DD channels are at work, Greggio's
(2005) study indicates that in early--type galaxies the DD channel
should prevail over SD events and in late--type galaxies a large
proportion of SNe~Ia may be from SD systems.

\section{CONCLUSION}

   Based on the above mentioned variety of single degenerate close
binary systems further mutli-wavelength long term monitoring and
modeling of the following promising SD candidates --- HR~8210
(=IK~Peg), HD~209295, V~471~Tau, U~Sco, RS~Oph, V394~CrA, CAL~83,
QR~And, RXJ0513.9-6951, V~Sagittae--type close binaries V~Sge,
V617~Sgr, T~Pyxidis and KPD~1930+2752 --- needs to be carried
out. These systems deserve high priority in observations and modeling
in order to better determine the masses and orbital parameters, to
further understand the accretion onto the white dwarfs, and to predict
whether the systems will become SNe~Ia.  

Because there are so many promising single--degenerate candidate
systems, and no known promising double--degnerates, we suspect that
SDs are responsible for some or perhaps all SNe~Ia, while DDs are
responsible for some or perhaps none.

This work has been supported by NSF grants AST-0204771 and
AST-0506028, and NASA LTSA grant NNG04GD36G.

\end{document}